\pdfoutput=1
\documentclass[]{raa}
\usepackage{graphicx,times}             %for PS/EPS graphics inclusion, new
\usepackage{natbib}
\usepackage{amssymb,amsmath}
\bibpunct{(}{)}{;}{a}{}{,}

\usepackage[pagebackref=true]{hyperref}

\usepackage{array}
\usepackage{graphicx,times}             %for PS/EPS graphics inclusion, new
\usepackage{natbib}
\usepackage{amssymb,amsmath}
\bibpunct{(}{)}{;}{a}{}{,}

\usepackage[pagebackref=true]{hyperref}

\usepackage{array}
\newcolumntype{$}{>{\global\let\currentrowstyle\relax}}
\newcolumntype{^}{>{\currentrowstyle}}
\newcommand{\rowstyle}[1]{\gdef\currentrowstyle{#1}%
  #1\ignorespaces
}
\usepackage[flushleft]{threeparttable}
\newcommand{\blockcomment}[1]{}
\blockcomment{
\newcommand{\apjl}{Astrophys. J. Lett.}   % Astrophysical Journal, Letters
\newcommand{\apjs}{Astrophys. J. Suppl. Ser.}   % Astrophysical Journal, Supplement
\newcommand{\ao}{Appl. Opt.}   % Applied Optics
\newcommand{\aap}{Astron. Astrophys.}   % Astronomy and Astrophysics
\newcommand{\aaps}{Astron. Astrophys. Suppl.}   % Astronomy and Astrophysics, Supplement
}

\renewcommand{\today}{2024-07-22}
\renewcommand{\timenow}{00:00}

\begin{document}

  \title{Near-Earth Asteroids as the Parents of the $\delta$-Cancrid Meteoroid Stream
}
%   \subtitle{I. Place Your Subtitle Here}

   \volnopage{Vol.0 (20xx) No.0, 000--000}      %%preserved for Editor. DOn't remove!
   \setcounter{page}{1}          %%starting page, preserved for Editor. DOn't remove!

   \author{G. I. Kokhirova
      \inst{\star1}
   \and M. Zhang %(周爱英) %% Put your Chinese name in "( )" if you like. Note to open line 11 "\usepackage[UTF8]{ctex}"
      \inst{\star2,3,4}
   \and X.-G. Li
      \inst{2,5}
   \and A. I. Zhonmuhammadi
      \inst{1}
   \and X. Liu
       \inst{2,3,4}
%   \and U. H. Khamroev
%      \inst{1}
%   \and M. N. Latipov
%      \inst{1}
\footnotetext[1]{The equal contributors and co-first authors of this paper.}
  }
%% Here is an example of three authors come from different institutes.
%% For single author or all the authors from an institute, use "\inst{}" only

   \institute{Institute of Astrophysics, National Academy of Sciences of Tajikistan, Ayni 299/5, Dushanbe, 734067, Tajikistan; {\it kokhirova2004@mail.ru}\\
%% Please give the E-mail address of the author, to whom future correspondence and
%% offprint requests will be sent.
        \and Xinjiang Astronomical Observatory, Chinese Academy of Sciences, 150 Science 1-Street, Urumqi 830011, China; {\it zhangm@xao.ac.cn}\\
        \and Key Laboratory for Radio Astronomy, Chinese Academy of Sciences, 2 West Beijing Road, Nanjing 210008, China\\
        \and Xinjiang Key Laboratory of Radio Astrophysics, 150 Science 1-Street, Urumqi 830011, China\\
        \and University of Chinese Academy of Sciences, 19A Yuquan Road, Beijing 100049, China\\
\vs\no
   {\small Received 20xx month day; accepted 20xx month day}}

\abstract{ The $\delta$-Cancrid meteoroid stream forms four active
  meteor showers which are observable on the Earth annually during
  January-February and August-September. The stream's definite parent
  comet has not been established. We performed a search for near-Earth
  asteroids (NEAs) associated with this stream. We have followed the
  backward evolution of the orbital elements of a sample of NEAs and
  found their orbits at the Earth-crossing positions. Using these
  orbits, we calculated the theoretical parameters of meteor showers
  associated with the considered NEAs. We carried out our search for
  observable active showers that match theoretically predicted ones
  with published data, and the result turned out that the predicted
  meteor showers of 13 NEAs were identified with the active showers
  produced by the $\delta$-Cancrid meteoroid stream. The comet-like
  orbits of NEAs and established association with active meteor
  showers indicate their common cometary origin. The NEAs considered
  are moving within the stream and likely represent the dormant
  remnants of a parent comet of the $\delta$-Cancrid
  asteroid-meteoroid complex disintegrated more than 12 thousand years
  ago.}  

   \authorrunning{G. I. Kokhirova \& M. Zhang et al.}            %author_head in even pages
   \titlerunning{$\delta$-Cancrid asteroid-meteoroid complex}  % title_head in odd pages

   \maketitle
%% The author head (on even pages) and the title head (on odd pages) will be
%% automatically extracted from \author{} and \title{}. Whenever the title is too long,
%% you will be asked to supply a shorter one by inserting either \authorrunning{} or
%% \titlerunning{} before \maketitle. Anyway, you can specify your own heads.
%%
%%
%% Note: In the following text body of your manuscript, please note several differences from
%%       other major journals:
%% (1) \subsection{Please Capitalize the First Letter of Each Notional Word in Subsection Title}
%% (2) Please Capitalize the First Letter of Each Notional Word in all tables' captions

%
%________________________________________________ sections below
%
\section{Introduction}           %% first-level sections will be auto-capitalized
\label{sect:intro}

Small bodies in the Solar System have archived the state of the
proto-solar disk throughout the Solar System's formation. Therefore,
studying small bodies in our Solar System will help us deeply
understand the formation and evolution of this unique planetary
system. There are a lot of minor bodies alongside with major planets
moving in our Solar System. The family of small bodies in our Solar
System includes comets, asteroids and meteoroids. In this hierarchy,
meteoroids are fragmental outputs of the disintegration of comets and
asteroids. A meteoroid component is divided into two main groups:
sporadic background and stream's meteoroids. In the context of our
paper the meteoroid stream will be considered so we give it definition
as following. A meteoroid stream is formed by a great number of
meteoroids generated by one parent body. The meteoroids belonging to
the same stream move in the interplanetary orbits close to the parent
body's orbit. It has been proven that the formation of meteoroid
streams can only be caused by the activity or destruction of
comets. As was shown by \citet{bredi1954}, only a periodic normal gas-
and dust- producing activity of a comet can form a stable and
long-lived meteoroid stream. This normal activity is observed when a
comet passes its orbital perihelion. In addition, a meteoroid stream
may result from the catastrophic disintegration of a comet as a result
of impact or other processes. The formation of a long-lived and
developed meteoroid stream cannot be ensured by a break-up or
so-called decay of an asteroid, such as a collision, since the
ejection of dust and debris will be one-time off and insufficient for
the formation of a stable stream in this case. The theory of the
meteoroid stream's formation due to cometary activity, the
circumstances of their evolution and structure are defined in a series
of papers (see e.g., \citealt{whipp1950, whipp1951, hughes1986,
  baba1992, baba2008mn, baba2015so}).

Of course, there are exceptions to this concept, the most famous and
well-studied of which is the connection between NEA (3200) Phaethon
and the Geminid meteor shower based on the similarity of orbits
\citep{whipp1983}, which is observed on Earth annually in the period
of December 10--15. There are a lot of papers showing that with a very
high probability Phaethon can be the parent body of the Geminid meteor
shower, on the basis of which the cometary nature of the asteroid is
assumed (see, for example, \citealt{will1993, ryab2019}). Moreover, in
2009, Phaeton was first recorded for the first time to have short-term
cometary activity in the perihelion region \citep{jewit2010} and then
this phenomenon was observed in 2012 and 2016 \citep{jewit2013,
  hui2017}, by this reason the object was classified as an active
asteroid \citep{jewit2012}. Phaeton's geometric albedo estimate of
0.107$\pm$0.011 (https://ssd.jpl.nasa.gov) corresponds to dark
asteroids and is consistent with an albedo range of 0.02--0.12 for
extinct cometary nuclei \citep{jewit1991}. On the other hand,
exploring the relationship between NEAs and showers, Wiegert, Brown
(2004) found that the link between the Geminids and Phaethon is
extremely unlikely and a mere chance alignment. Therefore, for a
convincing statement of the association between the primordial
asteroid and the meteor shower, both the connection with the shower
and the nature of Phaethon require further research.

As the orbit of the meteoroid stream intersects the Earth's orbit,
meteor showers are generated, which we can observe and record using
various techniques. As \citet{baba1992} showed, meteor streams,
depending on the number of intersections between their orbits and the
Earth’s orbit, can generate from four to eight observable annual
meteor showers on the Earth during the corresponding periods. However,
quadruple crossings are the most common case for meteoroid streams and
therefore the most typical production of a stream is four showers. To
demonstrate this, we use the example of the Taurid meteoroid stream,
which produces four meteor showers in a period. When the stream's
orbit intersects the Earth's orbit at the pre-perihelion, these are
the Northern and Southern Taurids, observed on the Earth annually from
September to November. At the post-perihelion crossing, these are the
Daytime $\beta$-Taurids and $\zeta$-Perseids, observed annually from
June to July. The parent body of Taurid stream is the comet 2P/Encke;
nevertheless, it turned out that more than 40 other near-Earth
asteroids (NEAs) belong to this family. A dynamic affinity of these
objects was established, and the family was named Taurid
asteroid-meteoroid complex. There is a very high probability that the
asteroids in the Taurid complex are in fact extinct comet nuclei or
dead fragments of a larger progenitor comet (see e.g.,
\citealt{asher1993, porub2006, baba2008mn}). The presence of a certain
number of extinct or dormant comet nuclei among the NEAs is beyond
doubt; according to some estimates, they may account for up to 6\% of
the total discovered NEAs currently \citep{opik1963, weiss1989, baba2012iau}.

According to \citet{weiss1989}, the term ``dormant'' or ``extinct''
comet refers to a comet nucleus that was active in the past and
currently loses its ability to generate a visible coma in any section
of its orbit. Throughout their evolution, such comet nuclei are
gradually covered with a thick and refractory mantle that prevents the
ejection of gas and dust; consequently, the normal cometary activity
ceases \citep{whipp1950, whipp1951, opik1963}. Meanwhile, an
extinct comet can be reactivated by a non-catastrophic collision with
another body or by bombardment of its surface by small meteoroids
\citep{weiss1989}. Such events have been confirmed by observations
\citep{baba2017aa}. According to ground-based observations, the
nuclei of extinct comets are indistinguishable from asteroids in
appearance. However, they can be distinguished by their dynamic
properties, i.e. their orbital elements. The typical cometary orbit
implies the cometary origin. The established connection between such
object and the observable active meteor showers will significantly
strengthen this assumption. Several asteroid-meteoroid complexes have
already been identified by using this approach, such as the Piscids
complex \citep{baba2008aa}, the $\iota$-Aquariids complex
\citep{baba2009}, the $\delta$-Scorpiids \citep{baba2013}, the
$\sigma$-Capricornids \citep{baba2015ad}, the Virginids
\citep{baba2012mn, baba2015aa,kokh2024}, etc. In addition to
meteoroid stream, each of these complexes contains several NEAs of
cometary origin, which may be the parents of relevant
streams. However, the parent bodies of all known meteoroid streams
have not yet been identified. Considering that finding the parents of
meteoroid streams is a critical step in understanding the genetic
connections between small Solar System objects, we continue to study
their relationships and discover new extinct comets in NEAs. This
paper presents the results of the discovery of relevant objects in the
$\delta$-Cancrid meteoroid stream.

%% Authors can give a citation as 'Michel et al. 1992'.
%% You may also use \cite, \citep and \citet for citation, and use Table~1 or Figure~1
%% and so forth. Using \ref and \label for cross-references of Tables/Figures
%% is a good way in adjusting/adding/removing text, tables or figures.

\section{$\delta$-Cancrid asteroid-meteoroid complex}
\label{sect:complex}

\subsection{Meteor showers of the $\delta$-Cancrid meteoroid stream}
\label{sect:stream}
In the meteor shower database of IAU MODC (www.ta3.sk, 2023), the
confirmed showers of the $\delta$-Cancrid stream are the nighttime
Northern and Southern $\delta$-Cancrids, 00096 NCC and 00097 SCC,
respectively, with a period of maximum activity at the end of January,
and the daytime southern shower Daytime $\chi$-Leonids having in the
database code 00204 DXL, with active period at the end of August. The
daytime northern shower has not been established. NEAs 1991 AQ and
2001 YB5 are indicated as possible parent bodies of the stream in the
IAU MODC database.

We searched for objects associated with this stream among NEAs
discovered before 2018, and identified 13 asteroids related to the
$\delta$-Cancrid stream, and also have established the northern branch
of the daytime shower and our results are presented hereinafter.

\subsection{Research approach and methodology}
\label{sect:method}
The research approaches and methods are based on the theory of
formation and evolution of meteoroid streams \citep{baba1992} and
the fact that there are a certain number of extinct comet nuclei among
NEAs \citep{opik1963, weiss1989, baba2012iau}. It turns out that
among the large number of meteoroids in the stream, only those with
orbital heliocentric distances equal to 1 AU for ascending R$_{\rm a}$
and descending R$_{\rm d}$ nodes can traverse the Earth's orbit
\citep{baba1992}. For most NEAs' orbits, this condition is
satisfied four times during one cycle of changing the orbital
perihelion argument. If the asteroid is indeed an extinct comet, then
a meteoroid stream could have been formed during the past cometary
activity. This meteoroid stream can theoretically cause four
observable meteor showers on the Earth -- the nighttime shower with the
northern and southern branches, as well as the daytime shower with the
northern and southern branches. To determine the parameters of these
theoretical showers, their orbits are needed. That is, the radiant,
velocity, and solar longitude of the predicted maximum activity for a
shower can be calculated using a set of orbital elements for the
asteroid at its Earth-crossing position.  This kind of asteroid orbit
that simultaneously corresponds to the theoretically predicted meteor
shower orbit can be obtained by calculating the orbital evolution of
the proposed parent body -- the asteroid. The orbital evolution is
calculated using various numerical integration methods of the
equations of motion for time intervals equal to one period of the
orbital perihelion argument change. As a rule, this time covers a
period of 10--12 thousand years for NEAs. The methods of
\citet{ever1974} and Halphen-\citet{gorya1937} are mostly used for the
orbital evolution calculation.

Once the characteristics of theoretical meteor showers have been
determined, accessible databases of observed meteor showers and
individual fireball/meteor showers should be searched for observable
activity that approximates the predicted showers. If the theoretical
meteor showers are consistent with observed meteor showers, then this
confirms a connection between the meteoroid streams that generate
these showers and the asteroids. In this case, it is quite possible to
assume that the asteroid have cometary properties.

In the end, only asteroids traveling in comet-like orbits and crossing
Earth's orbit were studied. Note that a comet-like orbit is the
necessary but not sufficient condition for an object's cometary
origin. We use the Tisserand parameter T$_{\rm j}$ to classify the
orbits here. If the Tisserand parameter value satisfies the condition
T$_{\rm j}\leq$3.12, then the orbit is classified as comet-like; and when
T$_{\rm j}\geq$3.12 the orbit is related to asteroidal type
\citep{kresak1969, jewit2012}. The condition for an asteroid's
orbit to intersect with the Earth's orbit have been verified in the
NEODyS-2 database (https://newton.spacedys.com, 2021).

\subsection{Near-Earth asteroids candidates for extinct comets}
\label{sect:candid}
In the NEOP database (http://www.neo.jpl.nasa.gov, 2019), among all
NEAs discovered till 2017-12-31 we selected asteroids which according
T$_{\rm j}$ are moving on the comet-like orbits. Then, among them we
selected NEAs which according the data of database NEODyS-2
(https://newton.spacedys.com, 2021) intersect the Earth's orbit (note,
not all NEAs orbits cross the Earth's orbit). By this way we got about
3 thousand asteroids. Then we calculated their orbital evolution
backward over the time interval equal to one cycle of the argument of
perihelion variations. As a result of calculation the evolution, a
number of NEAs ($\sim$3\%) were excluded due to chaotic motion. Using
the results of orbital evolution calculation for the remaining NEAs
the theoretical parameters of meteor showers were calculated. Finally,
we performed a computerized search for the observed
showers/meteors/fireballs/NEAs those parameters are close to the
theoretically predicted showers. Initially, the result has shown that
among approximately 2.5 thousand investigated NEAs theoretical showers
only of 13 asteroids were identified with the observed showers of the
$\delta$-Cancrid complex. Under this, in spite of identification of
the theoretical showers with the observed $\delta$-Cancrid showers
some NEAs were excluded since their values of $\pi$ are not consistent
with values of these 13 NEAs having averaged $\pi$=221$\pm$8$^\circ$ and of
the $\delta$-Cancrid showers having averaged $\pi$=220$\pm$10$^\circ$
(according to various published sources). This is an additional
condition that is met when selecting candidate asteroids for suggested
association. Note, as a result of a search, relations of some studied
asteroids from the sample to other new and known associations were
also determined which are the subject of future papers.

\begin{figure}
\centering
\includegraphics[width=\textwidth, angle=0]{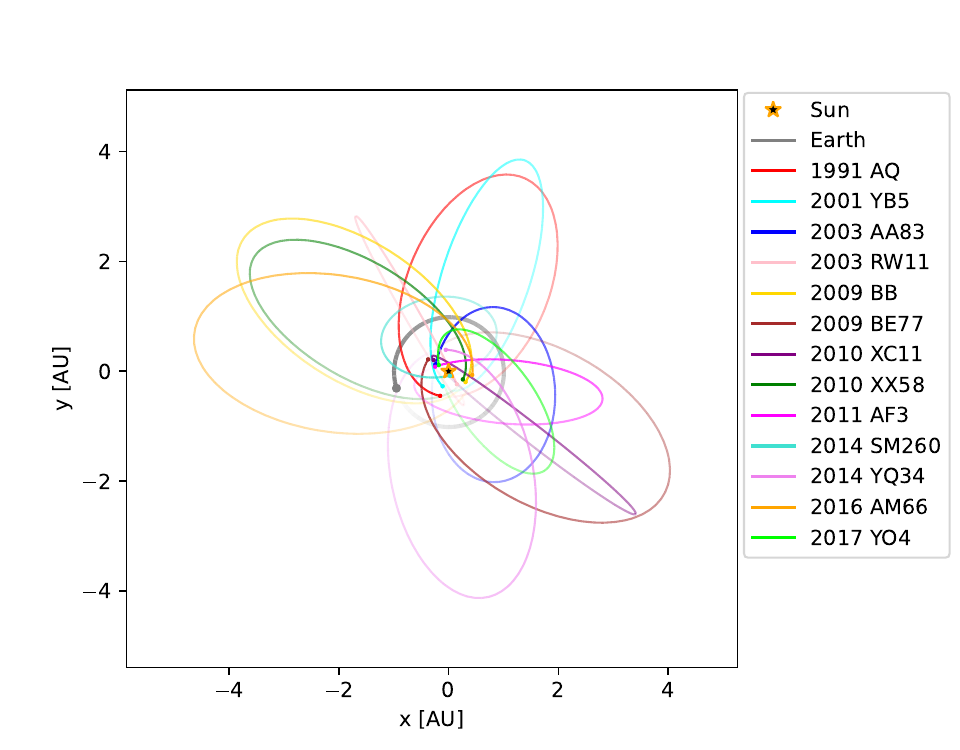}
\caption{Projected orbits of 13 NEAs on the ecliptic plane. The fading line of
  an orbit indicates the direction of the object's revolution.}
\label{eclip}
\end{figure}

The main parameters of 13 NEAS associated with $\delta$-Cancrid stream
are given in Table~\ref{Tab1}, including the asteroid designation, the
orbital elements (in Equinox 2000.0): a -- the semi-major axis, e --
the eccentricity, q -- the perihelion distance, i -- the inclination,
$\Omega$ -- the longitude of ascending node, $\omega$ -- the argument
of perihelion, $\pi$ -- the longitude of perihelion; as well as H --
the absolute magnitude, d -- the asteroid's equivalent diameter,
N$_{\rm i}$ -- the number of intersections of the asteroid's orbit
with the Earth's ones during one cycle of variation of the argument of
perihelion and T$_{\rm j}$ -- the value of Tisserand parameter, p --
geometrical albedo and S$_{\rm p}$ -- taxonomic classification. The
N$_{\rm i}$ value corresponds to the number of theoretically predicted
meteor showers associated with a given asteroid \citep{baba1992}. The
asteroid 1991 AQ has a value of T$_{\rm j}$=3.16, corresponding to the
boundary value of the criterion between comets and asteroids; however,
we classify the object's orbit as comet-like. Available information
about the albedo and taxonomy of three asteroids from the ALCDEF
database (www.alcdef.org) needs addition description. For NEA 1991 AQ
the albedo of 0.24$\pm$0.19 was measured with a large error; 2001YB5
has a measured albedo of 0.20, but the measurement error is not
given. In both cases, the albedo values are too uncertain; for this
reason, and taking into account the comet-like orbits, asteroids are
not excluded from consideration and a final conclusion on their nature
requests additional investigation. The albedo of 2003RW11 is measured
as 0.02$\pm$0.05 that is consistent with confirmed values for cometary
nuclei. These NEAs are spectrally classified as S-type, for which a
moderate albedo in the range 0.10--0.20 is typical
(www.alcdef.org). However, the albedo of 2003RW11 do not correspond to
the S-type, and given the uncertainties in the albedo values of 1991
AQ and 2001 YB5, it is not yet possible to make a conclusion about
their real spectral features. There are no data on the remaining 10
NEAs in the ALCDEF database, so for them we use the data accessible at
the ssd.jpl.nasa.gov database assuming their albedo to be relevant to
albedo values for cometary nuclei. For greater confidence and
strengthen the conclusions of our results, it is necessary to refine
NEAs albedo in the future. The diameters denoted by asterisk in
Table~\ref{Tab1} are accessible at the published databases such as
ssd.jpl.nasa.gov and alcdef.org. At the cases of a lack of size
estimation the diameters were calculated using the known expression
\citep{harris2002}
\begin{equation}\label{eq1}
  d = {1329 \over {\sqrt p \cdot 10^{0.2H}}},
\end{equation}
where p is the geometrical albedo of an asteroid. For very dark
asteroids of C, P, and D types, the albedo is as low as within
0.02--0.12, which indicates them very likely to be extinct comets
\citep{jewit1991}. Thus the quoted diameters of asteroids were
estimated with the median value of low albedos which is p=0.07. There
are 4 asteroids in this sample classified as potentially hazardous
ones in NEOP, and in Table~\ref{Tab1} they are denoted as PHA. The 2D
projection of the present orbits of 13 NEAs onto the ecliptic plane is
given in Fig.~\ref{eclip} where the Sun and the Earth's orbit are also
shown.

%
%               one-column-spanning table
%________________________________________ Table 1: Use_of_the routines
\begin{table}
\begin{center}
\caption[]{Main properties of NEAs related to the $\delta$-Cancrid complex. Asterisks indicate data sources.}\label{Tab1}
\resizebox{\textwidth}{!}{%
\begin{tabular}{lcccrrrrllcclc}
\hline\noalign{\smallskip}
Asteroid & a (AU) & e & q (AU) & i$^\circ$  & $\Omega^\circ$ & $\omega^\circ$  & $\pi^\circ$ & H (mag) & d (km) & N$_{\rm i}$ & T$_{\rm j}$ & p (\%) & S$_{\rm p}$ \\
\hline\noalign{\smallskip}
1991 AQ  (PHA) & 	2.221 &	0.777 &	0.494 &	3.1 &	339.5 &	241.1 &	222.6 &	17.2* &	1.1* &	4 &	3.16 & 0.24*$\pm$0.19 & S* \\
2001 YB5  (PHA)  & 	2.340 &	0.865 &	0.316 &	5.5 &	108.3 &	115.3 &	223.6 &	20.9* &	0.20* &	4 &	2.89 & 0.20*$\pm$? & S* \\
2003 AA83 &	2.452 &	0.773 &	0.555 &	6.8 &	87.7 &	127.4 &	215.1 &	21.8** &	0.21* &	4 &	2.98 & - & - \\
2003 RW11 &	2.635 &	0.824 &	0.465 &	10.3 & 	170.8 &	53.4 &	224.2 &	18.7* &	1.5* &	4 &	2.77 & 0.02*$\pm$0.05 & S* \\
2009 BB &	2.412 &	0.848 &	0.368 &	18.8 &	72.4 &	154.9 &	227.3 &	18.4** &	1.07 &	4 &	2.84 & - & - \\
2009 BE77 &	2.522 &	0.826 &	0.439 &	21.1 &	201.2 &	26.9 &	228.1 &	18.1** &	1.23 &	4 &	2.79 & - & - \\
2010 XC11 (PHA) &	2.516 &	0.850 &	0.377 &	9.1 &	94.3 &	121.2 &	215.5 &	18.7** &	0.94 &	4 &	2.79 & - & - \\
2010 XX58 &	2.238 &	0.864 &	0.305 &	22.2 &	53.1 &	174.6 &	227.7 &	18.6** &	0.95 &	4 &	2.94 & - & - \\
2011 AF3 &	2.312 &	0.817 &	0.422 &	7.8 &	99.5 &	112.2 &	211.7 &	25.1** &	0.05 &	4 &	3.01 & - & - \\
2014 SM260 (PHA) &	2.259 &	0.884 &	0.262 &	7.8 &	336.8 &	246.9 &	223.8 &	21.0** &	0.32 &	4 &	2.91 & - & - \\
2014 YQ34 &	2.480 &	0.827 &	0.428 &	3.4 &	272.3 &	290.6 &	202.9 &	24.1** &	0.08 &	4 &	2.87 & - & - \\
2016 AM66 &	2.558 &	0.830 &	0.435 &	19.7 &	235.6 &	355.0 &	230.5 &	20.0** &	0.51 &	4 &	2.77 & - & - \\
2017 YO4 &	2.237 &	0.829 &	0.382 &	7.4 &	189.6 &	26.8 &	216.4 & 20.6** &	0.33 &	4 &	3.05 & - & -\\
\hline\noalign{\smallskip}
\end{tabular}}
{\raggedright * https://alcdef.org;\quad ** https://ssd.jpl.nasa.gov/tools/sbdb\_lookup.html. \par}
\end{center}
\end{table}

\subsection{Investigation of orbital evolution}
\label{sect:orbit}
\begin{figure}
\centering
\includegraphics[width=0.49\textwidth, angle=0]{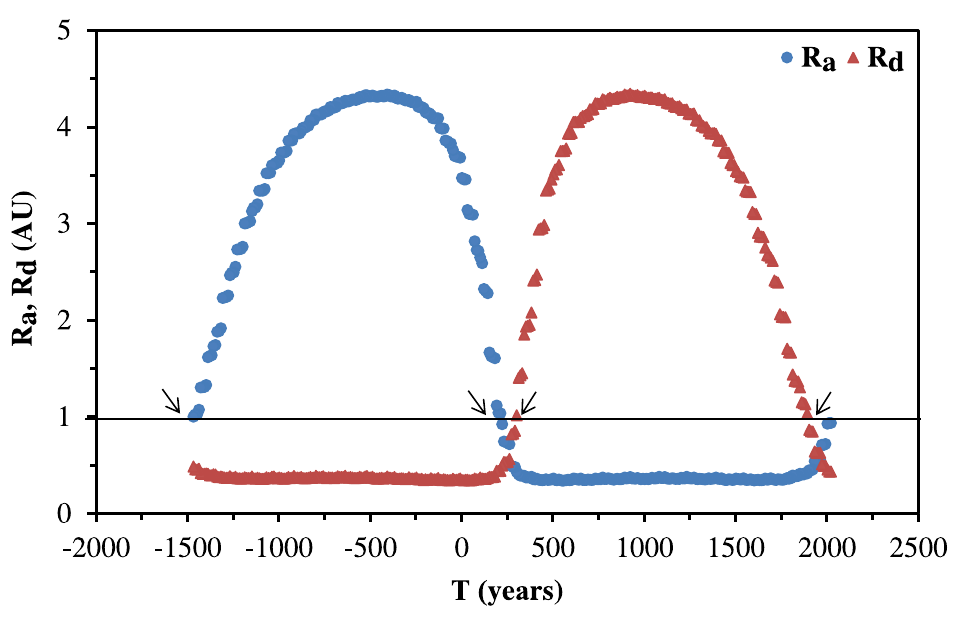}
\includegraphics[width=0.49\textwidth, angle=0]{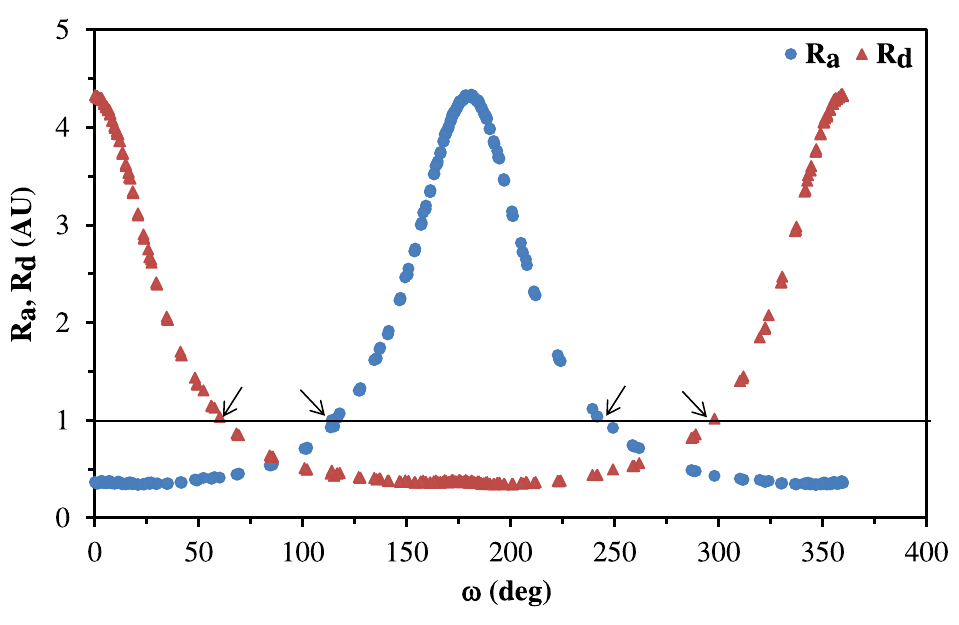}
\caption{Changes in R$_{\rm a}$ and R$_{\rm d}$ of the asteroid 2001 YB5's orbit with time (left) and argument of perihelion (right).}
\label{yb5}
\end{figure}

\begin{figure}
\centering
\includegraphics[width=0.49\textwidth, angle=0]{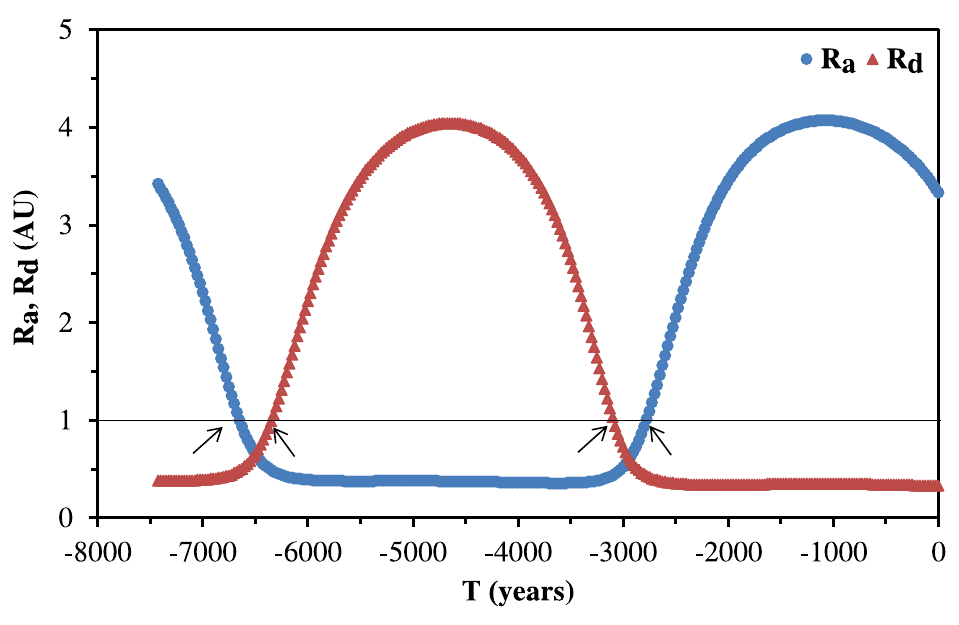}
\includegraphics[width=0.49\textwidth, angle=0]{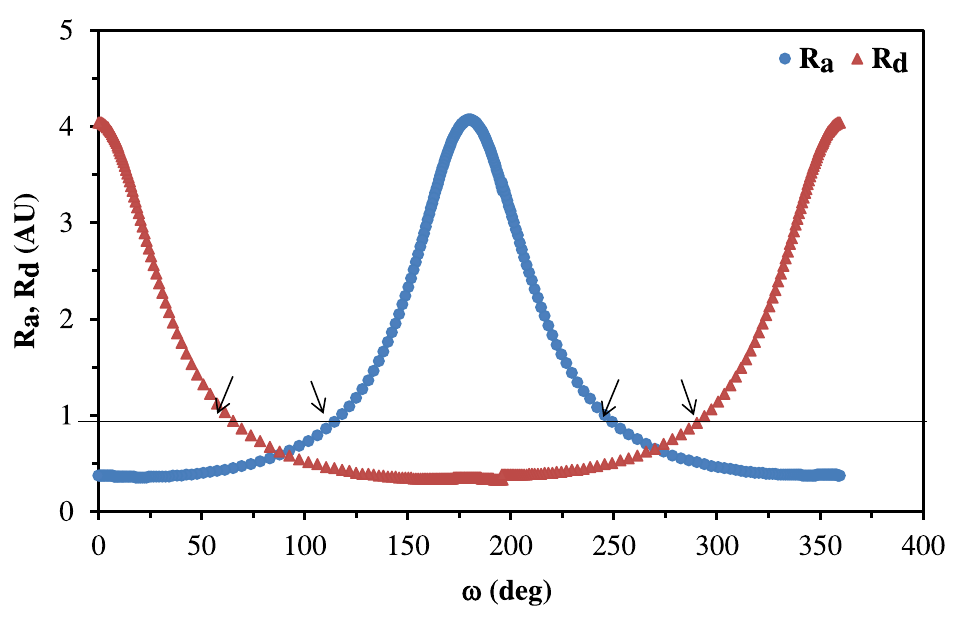}
\caption{Changes in R$_{\rm a}$ and R$_{\rm d}$ of the asteroid 2010 XC11's orbit with time (left) and argument of perihelion (right).}
\label{xc11}
\end{figure}

We calculated the orbital evolution of a sample of NEAs during a
periodic change in the argument of perihelion using the Everhart
RADAU19 method \citep{ever1974}. Then the gravitational
perturbations of the major planets were taken into account in the
evolution. As a result, it was found that all asteroids cross the
Earth's orbit four times during one cycle of $\omega$ variation. It
means that the sizes of R$_{\rm a}$ and R$_{\rm d}$ of asteroids
orbits equal to 1 AU four times in a cycle, i.e. twice at each
node. The variations of the R$_{\rm a}$ and R$_{\rm d}$ of the
asteroids orbits with time and the argument of perihelion are
demonstrated in Fig.~\ref{yb5}-\ref{xc11}, where the straight line
drawn parallel to the abscissa axis through the 1 AU value corresponds
to the position of the Earth's orbit, and the intersection with the
asteroid orbit is indicated by the arrows. These graphs are similar
for all considered NEAs, so we only present a sample of two plots for
a pair of asteroids here. For asteroid 2001 YB5, it is shown in
Fig.~\ref{yb5}; for asteroid 2010 XC11, it is shown in
Fig.~\ref{xc11}.

%%%%%%%%%%%%%%%%%%%%%%%%%%%%%%%%%%%%%%%%%%%%%%%%%%%%%%%%%%%%%%
%%     Examples for figures using graphicx for LaTeX 2e
%%               -- our recommended way for embodying graphics
%%%%%%%%%%%%%%%%%%%%%%%%%%%%%%%%%%%%%%%%%%%%%%%%%%%%%%%%%%%%%%
%
%      A figure as large as the width of the column
%-------------------------------------------------------------

The values of the argument of perihelion at the Earth's crossing
positions for the considered NEAs and averaged values with their
standard deviations are given in Table~\ref{Tab2}, where the
theoretical showers are listed in the following order: 1. the Northern
branch of nighttime shower, 2. the Southern branch of nighttime
shower, 3. the Northern branch of daytime shower, 4. the Southern
branch of daytime shower.

\begin{table}
\begin{center}
\caption[]{Values of the argument of perihelion of asteroids' orbits at the Earth's crossing positions. Averaged  values with their standard deviations are given in the last line.}\label{Tab2}
\begin{tabular}{lcccc}
\hline\noalign{\smallskip}
Showers	& Northern nighttime & Southern nighttime & Northern daytime & Southern daytime \\
\hline\noalign{\smallskip}
1991 AQ  (PHA)  & 278.2 &  97.5 & 83.7 & 262.1 \\
2001 YB5 (PHA)  & 297.2 & 114.2 & 61.8 & 244.2 \\
2003 AA83       & 262.6 &  90.1 & 88.1 & 275.6 \\
2003 RW11       & 280.6 & 100.8 & 80.3 & 260.2 \\
2009 BB	        & 290.2 & 114.9 & 68.3 & 249.9 \\
2009 BE77       & 295.6 & 115.2 & 67.6 & 247.6 \\
2010 XC11 (PHA) & 294.5 & 110.1 & 68.8 & 247.1 \\
2010 XX58       & 302.3 & 123.7 & 56.4 & 238.0 \\
2011 AF3        & 280.8 & 104.6 & 75.5 & 254.8 \\
2014 SM260 (PHA)& 303.9 & 123.8 & 56.2 & 235.3 \\
2014 YQ34       & 283.4 & 105.4 & 74.7 & 255.3 \\
2016 AM66       & 296.8 & 112.7 & 67.7 & 248.8 \\
2017 YO4        & 289.1 & 110.8 & 69.1 & 251.0 \\
Mean value of $\omega^\circ$ & 288.8$\pm$11.0 & 109.5$\pm$9.4 & 70.6$\pm$9.3 & 251.5$\pm$10.1 \\
\hline\noalign{\smallskip}
\end{tabular}
\end{center}
\end{table}

Using a set of orbital elements found from orbital evolution that
correspond to the intersection of the asteroid's orbit and the Earth's
orbit, we calculate the equatorial coordinates of the geocentric
radiants (i.e. the right ascension $\alpha_{\rm g }$ and the
declination $\delta_{\rm g}$), the geocentric velocities V$_{\rm g}$,
the solar longitudes L$_{\odot}$ and corresponding dates for
theoretical meteor shower activity associated with asteroids.

\subsection{Association of asteroids with observable showers of the $\delta$-Cancrid stream}
\label{sect:assoc}

A computer program to search theoretically predicted showers was
conducted in the published catalogs of the observable meteor/fireball
showers and detected fireball/meteors, meteorites.  By comparing
theoretical and observational parameters, we require that, the
difference between radiant positions should not exceed $\pm$10$^\circ$
in both right ascension and declination, the difference between
geocentric velocities should be $\Delta {\rm V}_{\rm g}\leq\pm$5 km/s
and the periods of activity may differ no more than $\pm$15 days (see,
e.g.,\citealt{baba2008aa,baba2008mn,baba2009,rudaw2015}). When these
conditions are fulfilled then we can declare a closeness of the
geocentric parameters of two compared showers-theoretical and observed
ones. The closeness of theoretical and observable orbits is checked
using the \citet{south1963} criterion D$_{\rm SH}$, which is a measure
of the similarity between two orbits. They also showed that the
threshold value of the D$_{\rm SH}$ criterion to confirm the
connection is 0.20. In this case, to calculate the threshold value
D$_{\rm SH}^{\rm max}$ while measuring similarity of the orbits of the
meteor and the shower, the following relation is used D$_{\rm SH}^{\rm
  max}$=0.2(360/N)$^{1/4}$, where N is the meteor data sample size. In
our case, where we compare 52 orbits (13 NEAs orbits * 4 multiple
intersections with the Earth's orbit=52) with the orbit of the active
shower, according to this expression D$_{\rm SH}^{\rm
  max}$=0.32. However, the use of a threshold D$_{\rm SH}$ criterion
value of 0.20--0.25 is generally accepted both when identifying streams
and when identifying the connection between the pairs a comet-shower,
an asteroid-shower, etc. Therefore, we settled on a value of
0.25. When the condition D$_{\rm SH}\leq$0.25 is met, the two orbits
are considered similar.

The search found a group of 13 NEAs (Table~\ref{Tab1}) closely
associated with meteoroid stream that produce nighttime Northern and
Southern $\delta$-Cancrids (96 NCC and 97 SCC) and Daytime Southern
$\chi$-Leonids (204 DXL). As mentioned above, the Northern branch of
the daytime shower has not yet been established. Our results suggest
that this may be the $\tau$-Cancrids meteor shower, listed as number
430 in the catalog of \citet{lebed1973}. The $\gamma$-Leonids meteor
shower from the \citet{sekan1976} catalog may also be the Northern
branch of the daytime shower, but its average orbit has a D$_{\rm SH}$
criterion value higher than that of the $\tau$-Cancrids shower, so we
chose the latter as the candidate of the Northern branch of the
daytime shower.

The search results for asteroids 1991 AQ and 2003 RW11 are shown in
Table~\ref{Tab3} and ~\ref{Tab4}. For the other asteroids in this set,
the results are the same. The meteor shower parameters, including
orbital elements, radiants, solar longitudes and activity dates of
theoretically predicted showers, are given in bold in Table~\ref{Tab3}
and ~\ref{Tab4}; the Northern branch of nighttime shower is labeled as
NNS, the Southern branch of nighttime shower is labeled as SNS, the
Northern branch of daytime shower is labeled as NDS, and the Southern
branch of daytime shower is labeled as SDS, where the letters N and D
mean the night- and day- time showers respectively. Catalogs in which
observable active showers and fireballs found identical to the
theoretical prediction are listed in Tables 3 and 4 as short codes: L1
\citep{lindblad1971}, S2 \citep{sekan1973}, S3
\citep{sekan1976}, J1 (\citep{jennis2007}, J
\citep{jennis2016},
%C \citep{cook1973}, 
R \citep{rudawska2014}, L \citep{lebed1973}, N \citep{nilsson1964},
MORP \citep{halli1996}, PN \citep{mccrosky1978}. As shown, the values
of the D$_{\rm SH}$ criterion confirm the similarity between the
theoretical and observable orbits. Identification of theoretical
meteor showers with observable meteor showers and fireballs is
confirmed by proximity of radiant position, velocity, and date of
activity. It then allows the conclusion that the connection between
the asteroids and these showers exists, and therefore, it is likely
that the NEAs under consideration are in fact of a cometary
origin. Furthermore, these objects may be considered as the parent
bodies of the stream, or they could be extinct fragments of a larger
comet precursor of the stream. This meteoroid stream just contains
large-sized remnants of the parent comet that is currently dormant.

\begin{table}
\begin{center}
\caption[]{Meteor showers, fireballs and NEAs related to NEA 1991 AQ (J2000.0).}\label{Tab3}
\resizebox{\textwidth}{!}{%
\begin{tabular}{$l^c^c^c^c^c^c^c^r^c^c^c^c^c^c}
\hline\noalign{\smallskip}
%Meteor showers, & a & e & q & i$^\circ$ & $\Omega^\circ$ & $\omega^\circ$ & L$_\odot^\circ$ & Date & $\alpha^\circ$ & $\delta^\circ$ & V$_{\rm g}$  & D$_{\rm SH}$ & Type & Catalog \\
Meteor showers, & $^1$a & $^2$e & $^3$q & $^4$i$^\circ$ & $^5\Omega^\circ$ & $^6\omega^\circ$ & $^7$L$_\odot^\circ$ & $^8$Date & $^9\alpha^\circ$ & $^{10}\delta^\circ$ & $^{11}$V$_{\rm g}$  & $^{12}$D$_{\rm SH}$ & $^{13}$Type & $^{14}$Catalog \\
%Meteor showers, & $\color{blue}^1$a & $\color{blue}^2$e & $\color{blue}^3$q & $\color{blue}^4$i$^\circ$ & $\color{blue}^5$$\Omega^\circ$ & $\color{blue}^6$$\omega^\circ$ & $\color{blue}^7$L$_\odot^\circ$ & $\color{blue}^8$Date & $\color{blue}^9$$\alpha^\circ$ & $\color{blue}^{10}$$\delta^\circ$ & $\color{blue}^{11}$V$_{\rm g}$  & $\color{blue}^{12}$D$_{\rm SH}$ & $\color{blue}^{13}$Type & $\color{blue}^{14}$Catalog \\
fireballs, NEAs & (AU) & & (AU) & & & & & & & & (km/s) & & & \\
\hline\noalign{\smallskip}
\rowstyle{\bfseries} NNS 1991AQ &	2.214 &	0.774 &	0.501 &	2.5 &	304.5 &	278.1 &	304.5 & 24 Jan & 131.8 & 21.0 &	24.5 &	- &	N &	- \\
00096  NCC &	2.273 &	0.803 &	0.448 &	0.3 &	297.1 &	282.9 &	297.1 &	17 Jan & 126.6 & 19.5 &	26.2 &	0.08 &	N &	L1 \\
00096  NCC &	1.901 &	0.777 &	0.425 &	1.2 &	292.9 &	287.9 &	292.9 &	13 Jan & 124.8 & 19.8 &	25.8 &	0.08 &	N &	S2 \\
00096  NCC &	1.829 &	0.783 &	0.397 &	1.5 &	296.4 & 291.3 &	296.4 &	16 Jan & 130.0 & 19.9 &	26.4 &	0.13 &	N &	S3 \\
00096  NCC &	2.230 &	0.814 &	0.410 &	2.7 &	290.0 &	286.6 &	290.0 &	10 Jan & 121.3 & 23.0 &	27.2 &	0.13 &	N &	J1 \\
%00096  NCC &	2.300 &	0.800 &	0.460 &	0.9 &	297.1 &	282.6 &	297.1 &	17 Jan & 125.5 & 20.4 &	25.9 &	0.18 &	N &	C \\
00096 NCC &	2.190 &	0.815 &	0.405 &	2.7 &	292.9 &	287.5 &	292.9 &	13 Jan & 125.2 & 22.0 &	27.3 &	0.13 &	N &	R \\
445 &	1.256 &	0.566 &	0.545 &	1.7 &	295.9 &	283.4 &	295.9 &	16 Jan &	125.3 &	22.1 &	18.0 &	0.22 &	N &	MORP \\
774 &	3.075 &	0.888 &	0.344 &	7.9 &	284.2 &	292.6 &	284.2 &	5 Jan &	119.6 &	27.0 &	31.2 &	0.16 &	N &	MORP \\
996 &	2.182 &	0.770 &	0.502 &	0.7 &	298.3 &	277.3 &	298.3 &	18 Jan &	124.2 &	20.6 &	24.4 &	0.10 &	N &	MORP \\
\rowstyle{\bfseries} SNS 1991AQ &	2.218 &	0.771 &	0.508 &	5.3 &	125.1 &	97.5 &	305.1 &	  25 Jan &	129.7 &	12.0 &	24.4 &	- &	N &	- \\
00097  SCC &	2.260 &	0.811 &	0.427 &	4.7 &	109.3 &	105.0 &	289.3 &	9 Jan &	118.5 &	16.1 &	26.9 &	0.15 &	N &	J \\
00097  SCC &	1.610 &	0.770 &	0.370 &	4.9 &	120.1 &	116.7 &	300.1 &	20 Jan & 133.9 & 12.6 &	26.3 &	0.24 &	N &	N \\
00097  SCC &	2.240 &	0.791 &	0.468 &	5.2 &	111.8 &	100.3 &	291.8 &	12 Jan & 118.8 & 15.1 &	25.7 &	0.15 &	N &	R \\
995 &	1.989 &	0.793 &	0.412 &	4.2 &	113.3 &	108.3 &	293.3 &	13 Jan &	124.0 &	15.6 &	26.6 &	0.10 &	N &	MORP \\
660110 & 1.960 & 0.790 &	0.412 &	3.9 &	109.6 &	107.5 &	289.6 &	10 Jan & 120.5 & 16.6 &	26.6 &	0.20 &	N &	PN \\
660113a & 1.330 & 0.670 &	0.439 &	8.4 &	112.7 &	111.7 &	292.7 &	13 Jan & 124.8 & 9.7 &	22.5 &	0.22 &	N &	PN \\
710131 & 3.120 & 0.830 &	0.530 &	9.1 &	130.7 &	91.6 &	310.7 &	31 Jan & 132.0 & 7.0 &	26.2 &	0.09 &	N &	PN \\
2001 YB5 &	2.349 &	0.862 &	0.324 &	5.5 &	109.4 &	114.2 &	289.4 &	10 Jan & 124.2 & 15.4 &	30.8 &	0.21 &	N &	- \\
\rowstyle{\bfseries} NDS 1991AQ &	2.227 &	0.768 &	0.517 &	5.3 &	138.9 &	83.7 &	138.9 &	12 Aug & 140.0 & 22.0 &	24.2 &	- &	D &	- \\
$\gamma$-Leonids &	1.969 &	0.710 &	0.571 &	7.0 &	152.5 &	87.3 &	152.5 &	26 Aug & 156.5 & 19.7 &	22.1 &	0.24 &	D &	S3 \\
430 ($\tau$-Cancrids) &	1.490 & 0.660 &	0.507 &	11.4 &	140.1 &	74.9 &	140.1 &	13 Aug & 139.1 & 30.7 &	21.9 &	0.18 &	D &	L \\
\rowstyle{\bfseries} SDS 1991AQ &	2.216 &	0.772 &	0.505 &	2.5 &	320.5 &	262.1 &	140.5 &	13 Aug & 138.2 & 13.1 &	24.5 &	- &	D &	- \\
00204 DXL &	1.598 &	0.793 &	0.330 &	2.5 &	334.9 &	238.4 &	154.9 &	28 Aug & 142.3 & 12.7 &	27.5 &	0.22 &	D &	S3 \\
2014 SM260 &	2.250 &	0.884 &	0.261 &	7.8 &	336.8 &	246.9 &	156.8 &	30 Aug & - & - & - &	0.25 &	D &	-\\
\hline\noalign{\smallskip}
\end{tabular}}
{\raggedright $^1$semi-major axis, $^2$eccentricity, $^3$perihelion
  distance, $^4$inclination, $^5$longitude of the ascending node,
  $^6$argument of perihelion, $^7$solar longitude, $^8$the date
  corresponding to the solar longitude, $^9$right ascension of the
  geocentric radiant, $^{10}$declination of the geocentric radiant,
  $^{11}$geocentric velocity, $^{12}$criterion of orbital similarity
  of \citet{south1963}, $^{13}$night or daytime- shower,
  $^{14}$published source. \par}
\end{center}
\end{table}

\begin{table}
\begin{center}
\caption[]{Meteor showers, fireballs and NEAs related to NEA 2003 RW11  (J2000.0).}\label{Tab4}
\resizebox{\textwidth}{!}{%
\begin{tabular}{$l^c^c^c^c^c^c^c^r^c^c^c^c^c^c}
\hline\noalign{\smallskip}
Meteor showers, & $^1$a & $^2$e & $^3$q & $^4$i$^\circ$ & $^5\Omega^\circ$ & $^6\omega^\circ$ & $^7$L$_\odot^\circ$ & $^8$Date & $^9\alpha^\circ$ & $^{10}\delta^\circ$ & $^{11}$V$_{\rm g}$  & $^{12}$D$_{\rm SH}$ & $^{13}$Type & $^{14}$Catalog \\
%Meteor showers, & $\color{blue}^1$a & $\color{blue}^2$e & $\color{blue}^3$q & $\color{blue}^4$i$^\circ$ & $\color{blue}^5$$\Omega^\circ$ & $\color{blue}^6$$\omega^\circ$ & $\color{blue}^7$L$_\odot^\circ$ & $\color{blue}^8$Date & $\color{blue}^9$$\alpha^\circ$ & $\color{blue}^{10}$$\delta^\circ$ & $\color{blue}^{11}$V$_{\rm g}$  & $\color{blue}^{12}$D$_{\rm SH}$ & $\color{blue}^{13}$Type & $\color{blue}^{14}$Catalog \\
fireballs, NEAs & (AU) & & (AU) & & & & & & & & (km/s) & & & \\
\hline\noalign{\smallskip}
\rowstyle{\bfseries} NNS 2003RW11 &	2.635 &	0.828 &	0.454 &	4.7 &	303.6 &	280.6 &	303.6 &	24 Jan & 133.4 & 22.5 &	27.0 & - & N &	- \\
00096  NCC &	2.273 &	0.803 &	0.448 &	0.3 &	297.1 &	282.9 &	297.1 &	17 Jan & 126.6 & 19.5 &	26.2 &	0.10 &	N &	L1 \\
00096  NCC &	1.901 &	0.777 &	0.425 &	1.2 &	292.9 & 287.9 &	292.9 &	13 Jan & 124.8 & 19.8 &	25.8 &	0.10 &	N &	S2 \\
00096  NCC &	1.829 &	0.783 &	0.397 &	1.5 & 	296.4 &	291.3 &	296.4 &	16 Jan & 130.0 & 19.9 &	26.4 &	0.10 &	N &	S3 \\
00096  NCC &	2.230 &	0.814 &	0.410 &	2.7 &	290.0 &	286.6 &	290.0 &	10 Jan & 121.3 & 23.0 &	27.2 &	0.12 &	N &	J1 \\
%00096  NCC &	2.300 &	0.800 &	0.460 &	0.9 &	297.1 &	282.6 & 297.1 &	17 Jan & 125.5 & 20.4 &	25.9 &	0.10 &	N &	C \\
00096 NCC &	2.190 &	0.815 &	0.405 &	2.7 &	292.9 &	287.5 &	292.9 &	13 Jan & 125.2 & 22.0 &	27.3 &	0.08 &	N &	R \\
774 &	3.075 &	0.888 &	0.344 &	7.9 &	284.2 &	292.6 &	284.2 &	5 Jan &	119.6 &	27.0 &	31.2 &	0.17 &	N &	MORP \\
996 &	2.182 &	0.770 &	0.502 &	0.7 &	298.3 &	277.3 &	298.3 &	18 Jan &	124.2 &	20.6 &	24.4 &	0.16 &	N &	MORP \\
\rowstyle{\bfseries} SNS 2003RW11 &	2.635 &	0.830 &	0.449 &	4.4 &	123.5 &	100.8 &	303.5 &	23 Jan & 130.9 & 13.5 &	27.2 &	- &	N &	- \\
00097  SCC &	2.260 &	0.811 &	0.427 &	4.7 &	109.3 &	105.0 &	289.3 &	9 Jan &	118.5 &	16.1 &	26.9 &	0.15 &	N &	J \\
00097  SCC &	1.610 &	0.770 &	0.370 &	4.9 &	120.1 &	116.7 &	300.1 &	20 Jan & 133.9 & 12.6 &	26.3 &	0.20 &	N &	N \\
00097  SCC &	2.240 &	0.791 &	0.468 &	5.2 &	111.8 &	100.3 &	291.8 &	12 Jan & 118.8 & 15.1 &	25.7 &	0.18 &	N &	R \\
995 &	1.989 &	0.793 &	0.412 &	4.2 &	113.3 &	108.3 &	293.3 &	13 Jan &	124.0 &	15.6 &	26.6 &	0.07 &	N &	MORP \\
660110 &	1.960 &	0.790 &	0.412 &	3.9 &	109.6 &	107.5 &	289.6 &	10 Jan & 120.5 & 16.6 &	26.6 &	0.12 &	N &	PN \\
660113a	 &1.330	& 0.670 & 	0.439 &	8.4 &	112.7 &	111.7 &	292.7 &	13 Jan & 124.8 & 9.7 &	22.5 &	0.18 &	N &	PN \\
710131 &	3.120 &	0.830 &	0.530 &	9.1 &	130.7 &	91.6 &	310.7 &	31 Jan & 132.0 & 7.0 &	26.2 &	0.12 &	N &	PN \\
2001 YB5 &	2.349 &	0.862 &	0.324 &	5.5 &	109.4 &	114.2 &	289.4 &	10 Jan & 124.2 & 15.4 &	30.8 &	0.13 &	N &	- \\
\rowstyle{\bfseries} NDS 2003RW11 &	2.635 &	0.830 &	0.448 &	4.1 &	144.0 &	80.3 &	144.0 &	17 Aug & 141.4 & 19.4 &	27.3 &	- &	D &	- \\
$\gamma$-Leonids  & 	1.969 &	0.710 &	0.571 &	7.0 &	152.5 &	87.3 &	152.5 &	26 Aug & 156.5 & 19.7 &	22.1 &	0.28 &	D &	S3 \\
430 ($\tau$-Cancrids) &	1.490 &	0.660 &	0.507 &	11.4 &	140.1 &	74.9 &	140.1 &	13 Aug &	139.1 &	30.7 &	21.9 &	0.25 &	D &	L\\
\rowstyle{\bfseries} SDS 2003RW11 &	2.635 &	0.828 &	0.454 &	4.4 &	324.1 &	260.2 &	144.1 &	17 Aug & 139.2 & 11.3 &	27.1 &	- &	D &	- \\
00204 DXL &	1.598 &	0.793 &	0.330 &	2.5 &	334.9 &	238.4 &	154.9 &	28 Aug & 142.3 & 12.7 &	27.5 &	0.20 &	D &	S3 \\
1991 AQ	 & 2.214 &	0.780 &	0.487 &	3.2 &	341.5 &	241.0 &	161.5 &	 4 Sep & - & - & - &	0.07 &	D &	- \\
2014 SM260 &	2.250 &	0.884 &	0.261 & 7.8 &	336.8 &	246.9 &	156.8 &	30 Aug & - & - & - &	0.21 &	D &	- \\
2014 YS43 &	2.960 &	0.823 &	0.524 &	12.9 &	326.8 &	244.9 &	146.8 &	19 Aug & - & - & - &	0.24 &	D &	- \\
\hline\noalign{\smallskip}
\end{tabular}}
{\raggedright $^1$semi-major axis, $^2$eccentricity, $^3$perihelion
  distance, $^4$inclination, $^5$longitude of the ascending node,
  $^6$argument of perihelion, $^7$solar longitude, $^8$the date
  corresponding to the solar longitude, $^9$right ascension of the
  geocentric radiant, $^{10}$declination of the geocentric radiant,
  $^{11}$geocentric velocity, $^{12}$criterion of orbital similarity
  of \citet{south1963}, $^{13}$night or daytime- shower,
  $^{14}$published source. \par}
\end{center}
\end{table}

In addition, the associated the NEAs listed in Table~\ref{Tab1}, with
theoretical and observable meteor showers and the connection between
each other, were found on the basis of the relevant radiants and the
current asteroids' orbits (see Table~\ref{Tab3}-\ref{Tab4}). A
proposed link between the stream and NEAs 1991 AQ and 2001 YB5
(www.ta3.sk, 2023) is also confirmed. As noted, in this family 4
asteroids are classified as potentially hazardous objects. The current
orbit of PHA 2001 YB5 corresponds to the Southern nighttime shower 97
SCC, while the current orbits of PHAs 2014 SM260 and 2014 YS43
corresponds to the Southern daytime shower
$\chi$-Leonids. Consequently, asteroids may enter the Earth's
atmosphere during associated meteor shower events and will have
characteristics similar to those of that shower. For instance, it is
theoretically predicted that the estimated impact date of the
potentially hazardous asteroid 2001 YB5 is January 10, the geocentric
speed is 30.8 kilometers/second, and the equatorial coordinates of the
point where it hits the Earth are $\alpha_{\rm g}$=$124.2^\circ$ and
$\delta_{\rm g}$=$15.4^\circ$.

\section{On the probability of random similarity of two orbits}
\label{sect:prob}

Since the proximity of orbits is not a sufficient condition to
confirm the connection between two objects moving in heliocentric
orbits, let us consider the probability of random similarity.

Assessing the plausibility of the associations between an NEA-type
object and a meteor shower is not a simple matter. Numerous attempts
were made to resolve this. For instance, \citet{wieg2004} have defined
expectation value of the number of asteroids closer to the shower
orbit than the test asteroid as P=N/(n), where N is the number of used
asteroids and n is the average number of trials to select asteroid's
orbits that have D$'$ criterion values satisfying condition D$'\leq$
D$'_0$ (D$'_0$ is value of the criterion of \citet{drumm1981} of the
test asteroid).  Further, they suggest that if this number is much
greater than one, then more than one asteroid is at least as well
aligned with the shower as the test asteroid, and so a chance
alignment becomes more probable. If this number is less than one, P
represents the probability that another asteroid is closer to the
shower than the chosen asteroid. A small value of P implies there are
few other asteroids in the phase space around the shower, and thus
that a chance alignment is unlikely. This approach was further
realized in \citet{ye2016}, where for five pairs asteroid-shower the
association was confirmed. Under this, the confirmation was supported
by the values of P around 1\%.

\citet{wieg2004} and further development of this approach reported in
\citet{ye2016}, when searching for the asteroid-shower linkage, they
initially used a random sample of a certain number of asteroids and
showers orbits. While, we used a slightly different approach; its
explanation and details on a selection procedure of objects for
consideration are given in section 2.3 from which it follows that the
sample of orbits was selected systematically, not random. It is
obviously that in our work were used to compare not all NEAs in the
JPLSSD database but only the 13 NEAs as candidates for the parent
bodies of the $\delta$-Cancrid complex found as a result of
systematical selection. By this reason, to assess the degree of
probability of a random coincidence of two orbits, we used the
methodology from our earlier work \citep{baba2008aa}. Such estimates
depend on the degree of similarity of the orbits. For clarity we
calculated the mutual D$_{\rm SH}$ criteria between the modern orbits
of first four studied NEAs (Table~\ref{Tab5}).

\begin{table}
\begin{center}
\caption[]{Mutual values of the D$_{\rm SH}$ criterion for NEAs.}\label{Tab5}
\begin{tabular}{lcccc}
\hline\noalign{\smallskip}
NEA	& 1991 AQ & 2001 YB5 & 2003 AA83  & 2003 RW11 \\
\hline\noalign{\smallskip}
1991 AQ   & 0    &  0.24 & 0.15 & 0.22 \\
2001 YB5  & 0.24 &  0    & 0.25 & 0.22 \\
2003 AA83 & 0.15 &  0.25 & 0    & 0.22 \\
2003 RW11 & 0.23 &  0.22 & 0.22 & 0 \\
\hline\noalign{\smallskip}
\end{tabular}
\end{center}
\end{table}

If asteroids were uniformly distributed in space, then the maximum
value of D$_{\rm SH}^2$ between extreme cases would be 5. However
asteroids inclinations are less than 30 degrees, so that for such a
set, the maximum value is 3 or 1.732 for D$_{\rm SH}$. Thus the
probability that two objects such as 1991 AQ and 2001 YB5, having
D$_{\rm SH}$ =0.24, by chance is 0.24/1.732 or about 14\%. For 1991 AQ
and 2003 AA83, where D$_{\rm SH}$ is 0.15, the probability that this
is by chance is about 9\%. The probability that all four are similar
by chance is thus about 3.5$\times10^{-4}$\%. For the associations
given in our work, the probabilities that these are random
associations is negligible. Thus it appears that, at a confidence
level $\sim$100\%, a subset of the selected NEAs is aligned with the
$\delta$-Cancrid complex. We recognize that the assumption of
uniformly distributed asteroids in space could be unjustified, and the
estimated probability of chance alignment could be off by an unknown
factor. However, taking into account a statement that, if the number
of asteroids having D$'\leq$D$'_0$ is large, the probability of a mere
chance association is high \citep{wieg2004}, we can consider this
assumption for the initial assessment of the probability of random
similarity of two orbits. Indeed, the fraction of asteroids in our
sample that satisfy this condition is $\sim$0.5\% of the total number
of asteroids, which also strengthens the assumption of a slight
probability of a random association. Finally, as was expressed by
\citet{wieg2004}, even if the probability of a chance association is
high, this does not exclude the existence of a real association
between the stream and the asteroid.

\section{A possible mechanism of the parent comet break-up}
\label{sect:breakup}

The fragmentation of asteroids and comets into large fragments occurs
at low ejection speeds. At the initial stage, small dispersion in the
elements of the orbits determines the low speeds at which the
fragments disperse. Therefore, when establishing the connection of
objects, it is necessary to study the evolution of orbits and find the
moment of their greatest similarity. If such similarity is found, then
this moment can be taken as the moment of separation of the fragments
(see, e.g., \citealp{khol2016, khol2017,baba2017aa, kokh2018}).
However, over time, due to various perturbations of both gravitational
and non-gravitational nature, the difference in the elements of the
two orbits can increase significantly, especially such angular
elements as $\Omega$ and $\omega$ (in this case, it is necessary to
control the stability of the value of $\pi$). For this reason, the
modern orbits of fragments originating from the same parent comet can
differ greatly from each other.  One short-coming in using the D$_{\rm
  SH}$ criterion for comparing orbits over long time intervals is that
the angular elements $\Omega$ and $\omega$ change over a reasonable
time scale so that D$_{\rm SH}$ can become large simply by these
changes. \citet{asher1993} proposed a simplified D criterion which
avoided this, namely
\begin{equation}\label{eq2}
  {\rm D}^2 = ({a_1-a_2 \over 3})^2 + (e_1-e_2)^2 + (2\sin{i_1-i_2 \over 2})^2,
\end{equation}
with the condition D$\leq$0.15 as showing similar orbits. 

\begin{figure}
\centering
\includegraphics[width=\textwidth, angle=0]{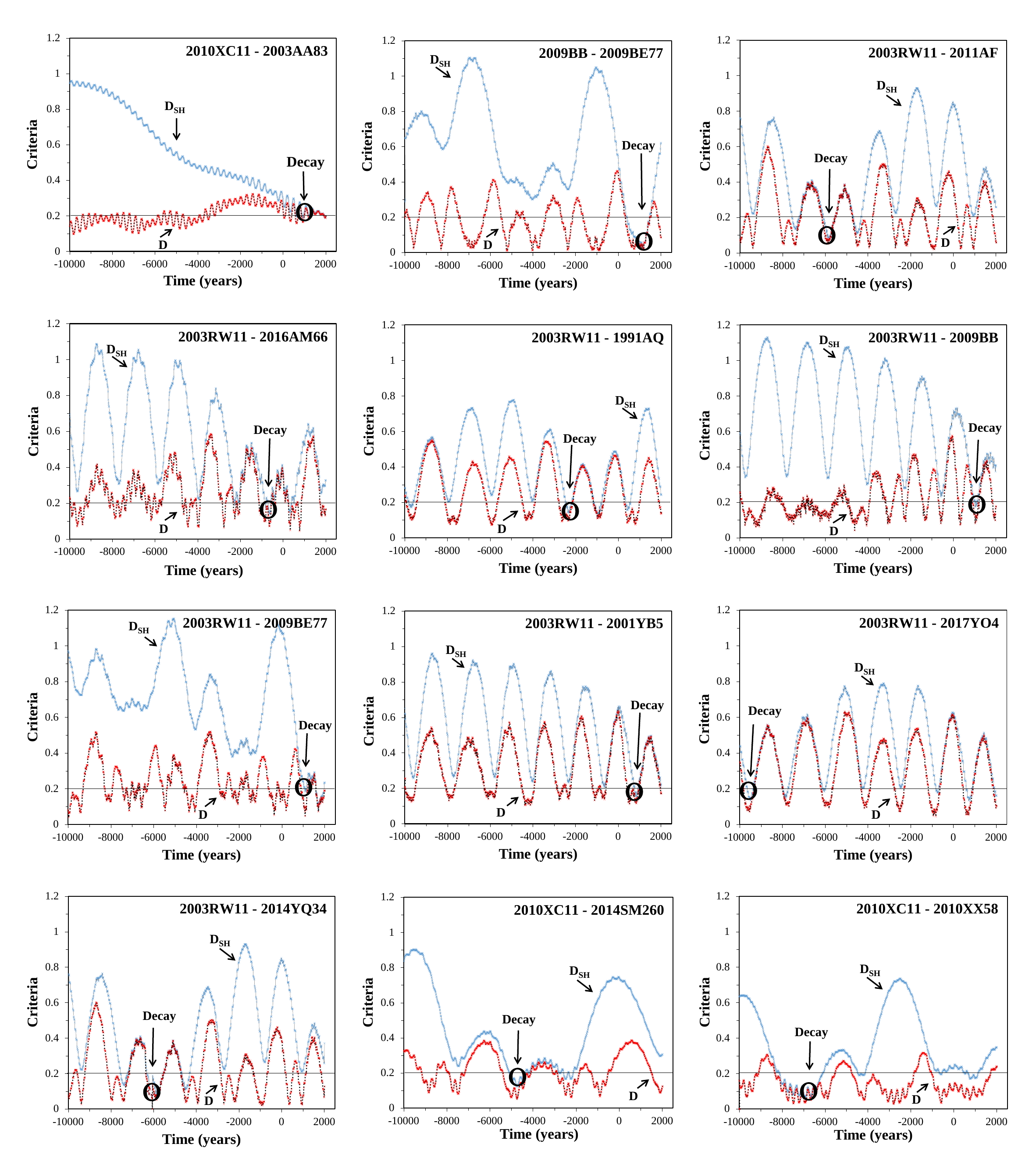}
\caption{Secular variations of the similarity criteria for NEAs
  fragmented from the parent comet of $\delta$-Cancrid complex. The
  abscissa plots the time interval studied, the ordinate gives the
  mutual value of the similarity criterion of the two orbits
  compared. The period of a decay is indicated by a big-O circle.}
\label{decay}
\end{figure}

In the absence of disturbances, the orbits will constantly pass
through the fragmentation point, that is, they will intersect at this
point. However, over time, this information disappears and is erased
due to various disturbances of both gravitational and
non-gravitational nature. To establish the moment of greatest orbital
similarity we calculated the D$_{\rm SH}$ criterion and simplified D
criterion of Asher et al (1993) between the orbits of studied NEAs and
have followed their behavior backward over 12 thousand years. The
secular variations of the both similarity criteria for the sample of
NEAs, between which fragmentaion very likely occurred, are presented
in Fig.~\ref{decay}. Note that we calculated both similarity criteria,
but in the analysis we relied mainly on the D$_{\rm SH}$ criterion as
the strongest indicator of the proximity of two orbits. Changes in
criterion D are presented to demonstrate their analogy to changes in
criterion D$_{\rm SH}$. In addition, the mutual values of the D
criterion for the studied NEAs, much lower than the accepted
threshold, along with the D$_{\rm SH}$ criterion, are an additional
indirect indicator of the similarity of the orbits of objects and,
therefore, their common origin. Analyzing obtained dependences we make
the following suggestions of a scenario for the parent comet break-up:

Starting from the four largest NEAs of our sample 2003 RW11 (1.5 km),
2009 BE77 (1.2 km), 2009 BB (1.1 km) and 1991 AQ (1.1 km) we
recognized that 2003 RW11 and 2009 BE77 have minimal values D$_{\rm
  SH}$=0.17 and D=0.05 around 1094 year; 2003 RW11 and 2009 BB have
minimal values D$_{\rm SH}$=0.15 and D=0.08 around 1094 year; 2009
BE77 and 2009 BB have minimal values D$_{\rm SH}$=0.02 and D= 0.02
around 1184 year. Fragmentation of 2009 BB and 2009 BE77 from 2003
RW11 have occurred $\sim$1094--1184 years, therefore about one thousand
years ago. 2003 RW11 and 1991 AQ have smallest values D$_{\rm
  SH}$=0.14 and D=0.11 around -2236 year, 1991 AQ probably broke away
from 2003 RW11 about 4.2 thousand years ago. Both pairs 2009 BE77 and
1991 AQ, 2009 BB and 1991 AQ do not have D$_{\rm SH}\leq$0.25 values
for the period under review, for this reason there was no
fragmentation between them.

Next, we examined the largest NEAs and medium-sized NEAs 2010 XC11
(0.9 km) and 2010 XX58 (0.9 km). It turned out that 2010 XC11 and 2010
XX58 have minimum values of D$_{\rm SH}$=0.07 and D=0.03 between
-6646 and -6966 years and it can be assumed that they separated
approximately 8.6--8.7 thousand years ago. 2010 XX58 and 2003 RW11 do
not have D$_{\rm SH}\leq$0.25 and D$\leq$0.15 values during considered
time, 2010 XC11 and 2003 RW11 have values of D$_{\rm SH}$ around
0.17--0.25 three times starting from 2014 till -2000 but they do not
correspond to smallest values of the D criterion. Consequently, 2003
RW11 could not have been fragmented into 20010 XX58 and 2010 XC11 that
confirms that the last two asteroids are debris of a single body. 2009
BE77 and 2010 XC11 do not have D$_{\rm SH}\leq$0.25 values, due to
this and taking into account the fact that 2009 BE77 appeared later
than 2010 XC, their split is impossible. The same was found for 2009
BE77 and 2010 XX58. This pair also does not have D$_{\rm SH}$ values
no bigger than 0.25 and given that 2009 BE77 occurred after 2010 XX58,
they could not have been fragmented. Although 2009 BB and 2010 XC11
have minimum values of D$_{\rm SH}$=0.08 and D= 0.01 between -3486 and -3526
years, however, since 2009 BB was allocated later and has lower
D$_{\rm SH}$ values with 2003RW and 2009 BE77 than with 2010 XC11, we
assume there was no division between them. Since 1991 AQ and 2010 XC11
do not have D$_{\rm SH}\leq$0.25 and D$\leq$0.15 values, and given
that 1991 AQ has separated much later than 2010 XC11 and 2010 XX58, a
break-up of 1991 AQ and 2010 XC11 is unlikely. The same thing is
observed for the pair of AQ 1991 and 2010 XX58. They also do not have
D$_{\rm SH}\leq$0.25 and D$\leq$0.15 values, and for the above reason,
the separation of 1991 AQ and 2010 XX58 could not occur.

Next, we included the remaining small-sized NEAs into
consideration. It is shown that 2003 AA83 (0.2 km) do not has D$_{\rm
  SH}\leq$0.25 values with 2003 RW11, 2009 BB, 2009 BE77, 2010 XX58
and 2014 YQ34, in addition, there are no D$\leq$0.15 values with 2010
XX58, which means that a fragmentation of the corresponding pairs
could not happen. However, 2003 AA83 and 2010 XC11 were found to have
minimum values of D$_{\rm SH}$=0.19 and D=0.16 between 964 and 914
years, leading to the conclusion that 2003 AA83 separated from 2010
XC11 about 1.1 thousand years ago. PHA 2001 YB5 has not values of the
D$_{\rm SH}$ criterion satisfying the threshold with 1991 AQ, 2009
BE77, 2014 SM260. The values of both criteria show that the orbit of
2001 YB5 (0.20 km) is similar with the orbits of 2009 BB, 2016 AM66
during almost total considered period, there are several minimal
values of both criteria between 2001 YB5 and 2010 XX58, 2010
XC11,etc. However, analyzing mutual values of both criteria for 2001
YB5 and 2003 RW11 we consider that their fragmentation very likely
occurred in the period of 864--844 years, about 1.2 thousand years ago.

2017 YO4 (0.33 km) has not values of the D$_{\rm SH}\leq$0.25 with
2009 BB, 2009 BE77, 2010 XX58, 2010 XC11, 2014 YQ34, 2003 AA83, and
2014 SM260 at studied period. 2003 RW11 and 2017 YO4 have smallest
values of D$_{\rm SH}$=0.13 and D=0.09 at the interval -9486 -- -9496
years. 2011 AF3 and 2017 YO4 have smallest values of D$_{\rm SH}$=0.13
and D=0.03 at the interval between -6386 and -6466 years. A closeness
of the orbits of 2017 YO4 and 1991 AQ is confirmed by the D criteria
values three times over considered period, and the orbits of 2017 YO4
and 2016 AM66 are similar four times, however, analyzing a behavior of
the mutual D criteria values we chosen a linkage of 2017 YO4 with the
largest 2003 RW11 as more likely.  Supposedly, 2017 YO4 was fragmented
from 2003 RW11 about 11.5 thousand years ago, and a closeness of the
2017 YO4 and 2011 AF3 orbits observed approximately 8.4--8.5 thousand
years ago is matching with a period of 2011 AF3 formation from 2003
RW11.

2014 SM260 (0.3 km) and 2010 XC11 have minimum values of D$_{\rm
  SH}$=0.14 and D=0.10 between -4656 and -4676 years, therefore it can
be assumed that 2014 SM260 has separated from 2010 XC11 almost 6.7
thousand years ago. 2014 SM260 and 2009 BE77 have minimum values of
D$_{\rm SH}$=0.14 and D=0.07 between -6416 and -6436, -6686 and -6696
years; and with 2009 BB it has minimum values of D$_{\rm SH}$=0.17 and
D=0.10 between -2236 and -2256 years, D$_{\rm SH}$=0.17 and D=0.11 at
-2436 and -2456 years. However, as established, 2009 BE77 and 2009 BB
appeared about one thousand years ago, and 2014 SM260 separated from
2010 XC11 6.7 thousand years ago. For this reason, despite satisfied
values of the criteria, 2014 SM260 could not separate from 2009 BE77
and 2009BB. The pairs 2014 SM260-2003 RW11 and 2014 SM260-2010 XX58 do
not have values of D$_{\rm SH}\leq$0.25 and D$\leq$0.15, so they are
likely not fragmented from each other. Because 2014 SM260 and 2003
AA83 have not values of the D$_{\rm SH}\leq$0.25 and D$\leq$0.15 and
taken into account that 2003 AA83 has divided from 2010 XC11 about 1.1
thousand years ago, a fragmentation of 2014 SM260 from 2003 AA83 is
impossible.

\begin{figure}
\centering
\includegraphics[width=0.8\textwidth, angle=0]{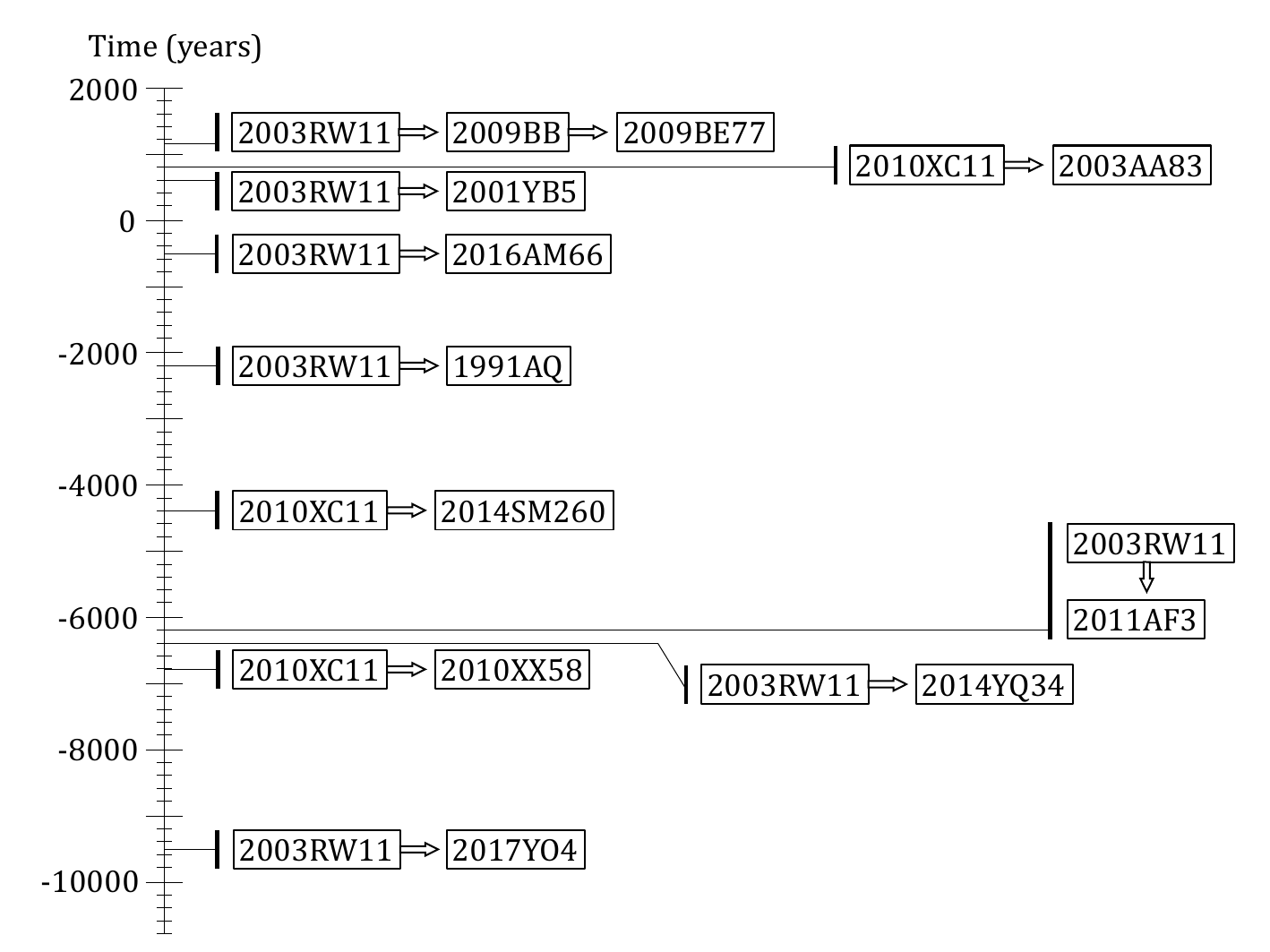}
\caption{Scheme of fragmentation of the parent comet of
  $\delta$-Cancrid complex. The time interval is plotted on the
  vertical axis and the boxes attached to the established decay period
  shows between which NEAs it occurred.}
\label{schem}
\end{figure}

According to the smallest values of D$_{\rm SH}$ and D criteria NEAs
2014 YQ34 (0.08 km) and 2011 AF3 (0.05 km) could separate
approximately between 7.8 and 8.1 thousand years ago. 2003 RW11 and
2014 YQ34 have minimal both criteria in -6096 year, therefore 2014
YQ34 separated from 2003 RW11 almost 8.1 thousand years ago. 2003 RW11
and 2011 AF3 have smallest both criteria in -5856 year and their
break-up occurred about 7.8 thousand years ago, consequently, 2011 AF3
and 2014 YQ34 fragmented from 2003 RW11 during 7.8--8.1 thousand years
ago. NEAs 2011 AF3 and 2016 AM66 (0.51 km) have minimal values of D
criteria in -8000 year, however, taking into account that 2016 AM66
was fragmented from 2003 RW11 2.6 thousand years ago while 2011 AF3
separated from 2003 RW11 7.8 thousand years ago, the pair 2011 AF3 and
2016 AM66 have not divided. NEA 2014 YQ34 has not values of the D${\rm
  SH}\leq$0.25 with NEAs 2009 BB, 2009 BE77, 2010 XC10, 2010 XX58,
2014 SM260, 1991 AQ so we can suggest there is not any fragmentation
between them. NEAs 2003 RW11 and 2016 AM66 have smallest values of
both criteria between -500 and -600 years, very probably, 2016 AM66
was separated from 2003 RW11 about 2.6 thousand years ago. The
proposed mechanism of parent comet disintegration is clearly shown in
the diagram in Fig.~\ref{schem}.

\section{Discussion}
\label{sect:discuss}

As a result of our study, a dynamic link between active meteor showers
generated by the $\delta$-Cancrid meteoroid stream and a set of
asteroids was established. This stream and the meteor showers it
generates, along with the 13 NEAs, together form the $\delta$-Cancrid
asteroid-meteoroid complex. The identified association between the
stream and showers is a very convincing indicator that 13 NEAs moving
in comet-like orbits have a cometary origin. Given the results of
dynamical modeling and following the behavior of the criteria of
orbital similarity we can assume the possible scenario of the
formation of this complex: initially, the giant parent Jupiter-family
comet of the stream was destructed into two large pieces more than 12
thousand years ago, there are 2003 RW11 and 2010 XC11. We can
confidently say that it was a giant comet, since almost half of its
remnants are on the order of or greater than 1 km in size. Next, the
cascade division of each of these objects began. 2017 YO4, 2011 AF3
and 2014 YQ34 were fragmented from 2003 RW11 about 11.5 and during the
period around 8 thousand years ago, correspondingly. 1991 AQ, 2016
AM66, 2001 YB5, 2009 BB and 2009 BE77 were separated from 2003 RW11
4.0--4.2, 2.5--2.6, 1.2, 1.0 and 0.8 thousand years ago,
correspondingly. 2010 XX58, 2014 SM260 and 2003 AA83 were fragmented
from 2010 XC11 8.9, 6.6--6.7 and 1.1 thousand years ago,
respectively. Of course, we realize that this is only a supposed
mechanism of fragmentation of the parent comet of the complex, and the
approach used for this assumption has certain shortcomings. For
greater persuasiveness of the given possible mechanism, it would be
necessary to statistically estimate the level of its probability
taking into account the number of discovered NEAs. Such a task is
beyond the scope of the present study and will be considered in the
authors' subsequent works. Thus, we can only suppose that objects were
formed by the disintegration of a giant comet -- the progenitor of the
stream and followed fragmentation of largest fragments. Presently, the
cometary objects are in an extinct stage. Coming from this we can
conclude that the $\delta$-Cancrid complex includes a meteoroid stream
that produces an observable active meteor shower and contains 13
large-sized extinct remnants of the parent comet. Furthermore, it was
revealed that in addition to small meteoroids, the stream also
contains large objects ranging in size from 50~meters to
1.5~kilometers. Among them, four asteroids are classified as
potentially hazardous objects.

\section{Conclusions}
\label{sect:conclu}

Our investigation suggests the establishment of a new $\delta$-Cancrid
asteroid-meteoroid complex. The complex includes the meteoroid stream
which produces active meteor showers and it is confirmed by the
observations. In addition, 13 near-Earth asteroids were found to be
probably dynamically associated with this complex. This association
and comet-like orbits indicate that the asteroids are likely of a
cometary origin. In this case, they may be regarded as the fragments
of a parent comet of the $\delta$-Cancrid complex, currently in an
extinct or dormant phase. We should acknowledge the advanced nature of
the method introduced by \citet{ye2016}, especially with the
development of computing technology today. Our analytical method uses
a simplified assumption based on a concept of formation and evolution
of meteoroid streams, following and comparing the objects' orbital
evolution. However, this widely used approach has yield the
establishment of a set of known complexes, whose reliability has been
confirmed by numerous publications including \citet{wieg2004} and
\citet{ye2016}. Under-estimation the probability of chance association
is the main shortcoming of the assumption. It is always a challenge to
distinguish a genuine parent-stream linkage from a chance alignment,
which is further complicated by the difficulty in obtaining their
precise orbits and fragmentation history \citep{ye2022}. However, it
is still a convenient approach to get the right probability at least
to the order of magnitude. For more realistic estimation of the
association between the meteor showers and asteroid orbits, we need to
utilize Monte Carlo method to do Bayesian prediction with objective
priors, suppressing the selection bias, as described by
\citet{ye2016}. The proposed scenario of cascade fragmentation of the
cometary parent is rather a phenomenological interpretation than a
physical mechanism so that it can be considered only as possible
mechanism. For more physical inference, we should introduce the
physical criterion of asteroid fragmentation and do the Monte Carlo
simulation to get the possible physical forms and probability of
fragmentation.  The results point to that meteoroid streams contain
both small particles on the millimeter scale and large objects over
the decameter scale, posing a potential hazard to the Earth and space
exploration missions. This is confirmed by both theoretical studies
and observational data.

Our research allow prediction of parameters for such objects entering
the Earth's atmosphere, which is needed to develop mitigation
strategies to prevent the consequences. The search for other putative
large fragments of the parent comet should continue in the
future. They are also extinct at present and may be found in known
NEAs and newly discovered NEAs.

\begin{acknowledgements}
The authors express their deep gratitude to the anonymous reviewers
for their careful study and prudent comments on the article, which
greatly improved the scientific level of the work. The N-body
integrator REBOUND \citep{rein2015} was utilized to model and plot the
asteroids' orbits in this paper. Co-authors from China acknowledge the
support from the National Key R\&D Intergovernmental Cooperation
Program (2023YFE0102300/2022YFE0133700), the Regional Collaborative
Innovation Project of Xinjiang (2022E01013), the National Natural
Science Foundation (12173078) and the "Belt and Road" Innovative
Talent Exchange Program (DL2023046004).
\end{acknowledgements}

%\appendix                  %%appendicial material is supported
%\section{This shows the use of appendix}

\bibliographystyle{raa}
\bibliography{meteor}
\label{lastpage}

\end{document}